\documentclass[sigplan]{acmart}  
\pdfoutput=1

\renewcommand\footnotetextcopyrightpermission[1]{} 

\AtBeginDocument{%
  \providecommand\BibTeX{{%
    \normalfont B\kern-0.5em{\scshape i\kern-0.25em b}\kern-0.8em\TeX}}}

\usepackage{enumerate}
\usepackage{romannum}
\usepackage{amsmath}
\usepackage{pdflscape}
\usepackage{booktabs} 
\usepackage{amssymb}
\usepackage{caption}

\usepackage{listings}
\usepackage{color}

\usepackage[english]{babel}
\usepackage{blindtext}

\usepackage{graphicx}
\usepackage{comment}

\usepackage{etoolbox}
\makeatletter
\patchcmd{\@makecaption}
  {\scshape}
  {}
  {}
  {}
\makeatother
\usepackage{threeparttable}
\usepackage{tablefootnote}
\usepackage{footnote}
\usepackage{fancyhdr}
\makesavenoteenv{tabular}
\makesavenoteenv{table}
\usepackage{eqparbox}
\usepackage{color, colortbl}
\usepackage{multirow}
\usepackage[linesnumbered,ruled,norelsize]{algorithm2e}

\usepackage{booktabs} 

\usepackage{color}
\usepackage{microtype}
\usepackage[colorinlistoftodos]{todonotes}
\usepackage{mathtools}
\usepackage{mdframed}
\usepackage{listings}
\usepackage{color}
\usepackage{url}
\usepackage{balance}
\usepackage{comment}
\usepackage{cancel}
\usepackage{array}
\newcolumntype{L}[1]{>{\raggedright\let\newline\\\arraybackslash\hspace{0pt}}m{#1}}
\newcolumntype{C}[1]{>{\centering\let\newline\\\arraybackslash\hspace{0pt}}m{#1}}
\newcolumntype{R}[1]{>{\raggedleft\let\newline\\\arraybackslash\hspace{0pt}}m{#1}}

\newcommand{\remove}[1]{\iffalse\sout{#1}\fi}

\newcolumntype{L}[1]{>{\raggedright\let\newline\\\arraybackslash\hspace{0pt}}m{#1}}
\newcolumntype{C}[1]{>{\centering\let\newline\\\arraybackslash\hspace{0pt}}m{#1}}
\newcolumntype{R}[1]{>{\raggedleft\let\newline\\\arraybackslash\hspace{0pt}}m{#1}}

\usepackage{graphicx}
\usepackage{subfig}
\usepackage[linesnumbered,ruled,norelsize]{algorithm2e}
\usepackage{color}
\usepackage{algpseudocode}

\usepackage{stfloats}
\usepackage{multirow}
\usepackage{todonotes}
\definecolor{commentgreen}{RGB}{2,112,10}

\SetCommentSty{mycommfont}
\makeatletter
\newcommand{\removelatexerror}{\let\@latex@error\@gobble}

\makeatother

\begin{document}
	
\author{{Bangtian Liu}}
\authornote{Equal contributions}
\affiliation{%
	\department{\textit{CS Department}}
	\institution{University of Toronto}
	\city{Toronto}
	\country{Canada}
}
\email{bangtian@cs.toronto.edu}
\author{{Kazem Cheshmi}}
\authornotemark[1]
\affiliation{%
	\department{\textit{CS Department}}
	\institution{University of Toronto}
	\city{Toronto}
	\country{Canada}
}
\email{kazem@cs.toronto.edu}
\author{ { Saeed Soori}}
\affiliation{%
	\department{\textit{CS Department}}
	\institution{University of Toronto}
	\state{Toronto}
	\country{Canada}
}
\email{sasoori@cs.toronto.edu }
\author{Michelle Mills Strout}
\affiliation{%
	\department{\textit{CS Department}}
	\institution{University of Arizona}
	\city{Tucson}
	\country{USA}
}
\email{mstrout@cs.arizona.edu}
\author{{ Maryam Mehri Dehnavi}}
\affiliation{%
	\department{\textit{CS Department}}
	\institution{University of Toronto}
	\city{Toronto}
	\country{Canada}
}
\email{mmehride@cs.toronto.edu}

\title{MatRox:  Modular approach for improving data locality in  Hierarchical (Mat)rix App(Rox)imation}


\begin{abstract}
Hierarchical matrix approximations have gained significant traction in the machine learning and scientific community as they exploit available low-rank structures in kernel methods to compress the kernel matrix. 
The resulting compressed matrix,  \textit{HMatrix}, is used to reduce the computational complexity of operations such as HMatrix-matrix multiplications with tuneable accuracy in an \textit{evaluation} phase.  
Existing implementations of HMatrix evaluations do not preserve locality and often lead to unbalanced parallel execution with high synchronization. Also, current solutions require the compression phase to re-execute if the kernel method or the required accuracy change.  
In this work, we describe MatRox, a framework that uses novel structure analysis strategies, blocking and coarsen, with code specialization and a storage format  to improve locality and create load-balanced parallel tasks for HMatrix-matrix multiplications.  
Modularization of the matrix compression phase enables the reuse of computations when there are changes to the input accuracy and the kernel function.   The MatRox-generated code for matrix-matrix multiplication is $2.98\times$, $1.60\times$, and $5.98\times$  faster than library implementations available in GOFMM, SMASH, and STRUMPACK respectively.  Additionally, the ability to reuse portions of the compression computation for changes to the accuracy leads to up to $2.64\times$ improvement with MatRox over five changes to accuracy using GOFMM.

\end{abstract}

\maketitle

\section{Introduction}

A large class of applications in machine learning and scientific computing involve computations on a dense symmetric positive definite (SPD) matrix that is  obtained by  computing a kernel function $\mathcal{K}$ on pairs of points from a set of points $\{x_1, \ldots, x_N\}$. The values of the  $N \times N$ kernel matrix $K$ are given by $ K(i,j) = \mathcal{K}(x_i, x_j)$ with a typically large $N$.   For example, in Gaussian ridge regression, 
the kernel $exp(- || x_i - x_j ||_2^2/2h^2)$,  $h$ is bandwidth,  is applied to a machine learning dataset, i.e. points. The resulting kernel matrix is used in costly matrix-matrix  multiplications, with complexity $O(N^3)$,   in a direct solver to minimize a loss function. 

The computational complexity of kernel matrix computations is reduced significantly, leading to orders of magnitude performance \cite{borm2007approximating}, if instead of assembling $K$ and operating on it, it is compressed to $\widetilde{K}$ using the kernel function,  points, and an \textit{admissibility condition} \cite{borm2003introduction,hackbusch2004hierarchical}.  An admissibility condition is the value
of a distance measure between points above with which 
the kernel value for that point pair is approximated.

Many of kernel matrices are \textit{structured} (or \textit{low-rank}, or \textit{data-sparse}). Hierarchical matrix computations, leverage the
data-sparse structure induced by the point set distribution and admissibility condition during a \textit{compression} phase to implicitly obtain $\widetilde{K}$.  First a \textit{cluster tree (CTree)} is created from a partitioning of points. Compression then uses an \textit{HTree}, a CTree which includes interactions from the admissibility condition, to hierarchically approximate low-rank blocks of the kernel matrix.  The low-rank approximated blocks as well as the blocks that are not approximated are referred to as \textit{submatrices} and form the  compressed matrix  $\widetilde{K}$ also known as the \textit{HMatrix}. The submatrices are then  operated on instead of $K$ in an \textit{evaluation} phase.

Previous work attempts to optimize hierarchical matrix computations, specifically the evaluation phase,  on parallel multicore architectures~\cite{yu2017geometry,ghysels2016efficient,kriemann2005parallel}. $\mathcal{H}^2$ structures are amongst the most commonly used hierarchical algorithms and GOFMM~\cite{yu2017geometry}, STRUMPACK~\cite{ghysels2016efficient}, and SMASH~\cite{cai2018smash} are well-known 
libraries that implement $\mathcal{H}^2$ structures. 
However, these  implementations do not preserve locality and often lead to a load-imbalanced execution with high synchronization overheads, which limits the performance and scalability of hierarchical matrix evaluations on parallel architectures.  


The order and dependency of computations during evaluation is determined by the HTree. GOFMM~\cite{yu2017geometry} uses the HTree as the input for dynamic task scheduling, however, their scheduling trades locality for load balance. SMASH~\cite{cai2018smash} traverses the CTree level-by-level, thus, synchronization overheads increase with the length of the critical path. Also schedulers that work with the CTree do not realize the additional dependencies introduced by the admissibility condition, which leads to additional synchronization costs.  Implementations such as SMASH and STRUMPACK~\cite{ghysels2016efficient} do not optimize for load balance. 
Finally, some libraries are optimized for a specific $\mathcal{H}^2$ structure, for example STRUMPACK is specialized for Hierarchical Semi-separable (HSS)~\cite{chandrasekaran2006fast} structures, or only support low-dimensional points, e.g. SMASH.

In this paper, we present structure analysis algorithms based on a modularized compression to generate code that improves data locality and maintains a good load balance for HMatrix evaluations. Our work focuses on HMatrix-matrix multiplications for the evaluation phase; we use the words HMatrix evaluation and HMatrix-matrix multiplication interchangeably throughout the paper. Our structure analysis uses a novel blocking algorithm and a coarsen algorithm based on load-balanced level coarsening (LBC) \cite{cheshmi2018parsy} to generate specialized code for evaluation and to store the submatrices in the order that they are visited during evaluation.

The proposed algorithms are implemented in a framework called MatRox,
which uses structure information from the points, the kernel function, as well as the  admissibility and accuracy requirement. The MatRox \textit{inspector} compresses $K$, analyzes structure, and generates optimized code. The \textit{executor} computes the  HMatrix-matrix multiplication.  MatRox supports  $\mathcal{H}^2$ using a binary cluster tree 
and takes as input low- and high-dimensional points.

Additionally, MatRox enables partial reuse of computations when the  kernel function and/or input accuracy are modified. 
In scientific and machine learning simulations, often the kernel matrix has to be re-compressed because either the overall accuracy of the HMatrix-matrix multiplication  is not sufficient or has to be reduced for faster evaluation, or the kernel function changes. While available libraries have to rerun the costly compression, MatrRox reuses parts of the previous inspection, e.g. compression information, and also reuses the previously generated evaluation code. 

The main contributions of this work include:
\begin{itemize}
\item Two novel structure analysis strategies, based on the modularization of compression, called blocking and coarsen, that enable the generation of specialized code for HMatrix-matrix multiplications. The specialized code maintains an efficient trade-off between locality, load balance, and parallelism.

\item The Compressed Data-Sparse (CDS) storage format  that follows the data access pattern during HMatrix evaluation to improve  locality.

\item Implementation of the proposed strategies in a framework called MatRox. The MatRox-generated  code is on average $2.98\times$, $1.60\times$, and $5.98\times$  faster than GOFMM, SMASH, and STRUMPACK respectively. 

\item An approach  that enables the reuse of computations in compression for when the kernel function and/or  accuracy change. MatRox with reuse is  $2.21\times$ faster than GOFMM over 5 changes to the input accuracy.
\end{itemize}

\section{Approach Overview}

In this section, we review the typical approach to hierarchical approximation and then provide an overview of the approach presented in this paper that is implemented in MatRox.

\begin{figure*}
\centering
\begin{tabular}{lll}

 \subfloat[Example pointset that has been hierarchically partitioned.  For example partitions 3 and 4 are both included in partition 1.]{
 \label{fig:points}
 \includegraphics[width=0.28\textwidth]{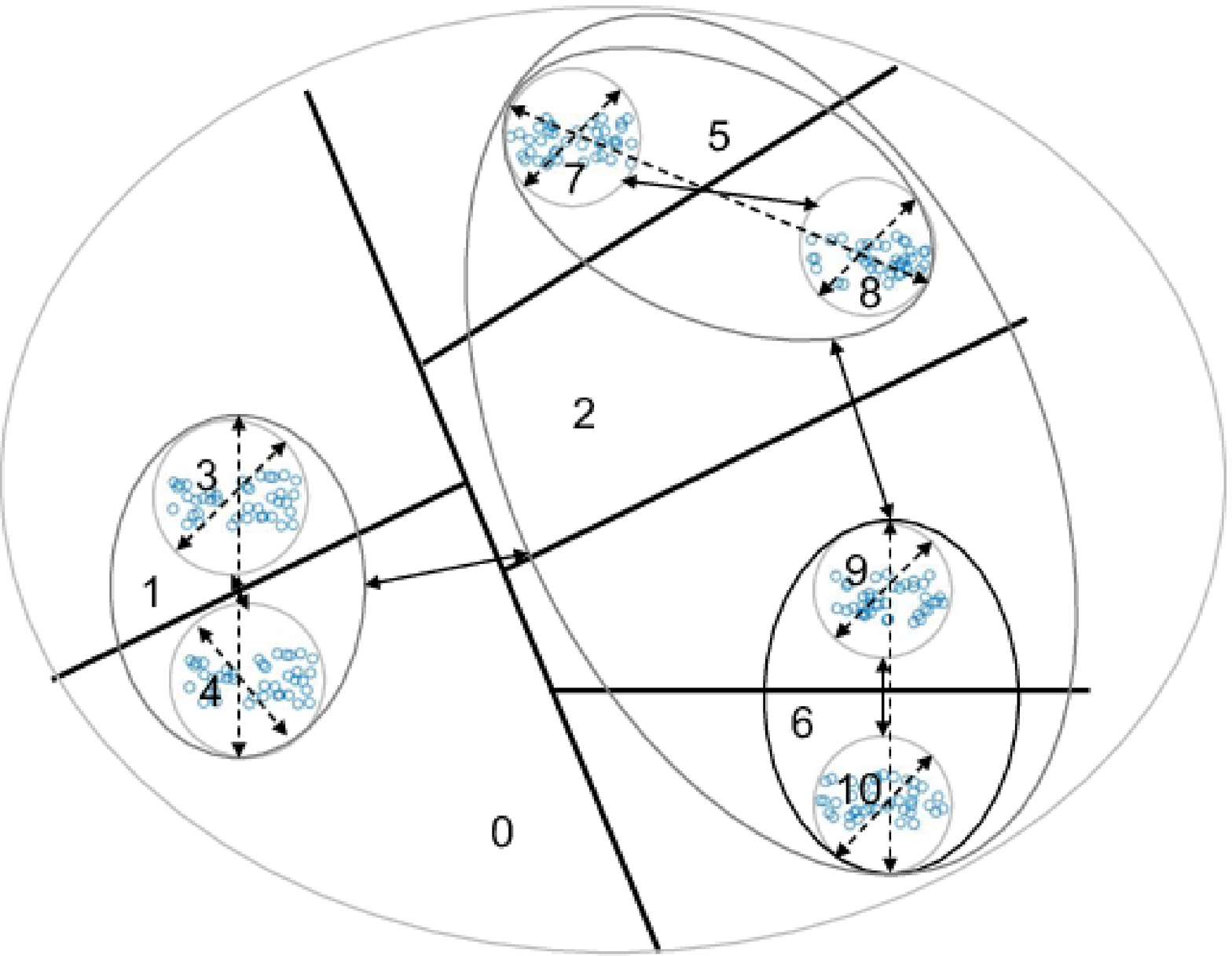}
 }    & 
 \subfloat[Cluster tree and Htree.  Each node represents a partition.  Edges between nodes indicate an interaction.]{
 \label{fig:htree}
 \includegraphics[width=0.33\textwidth]{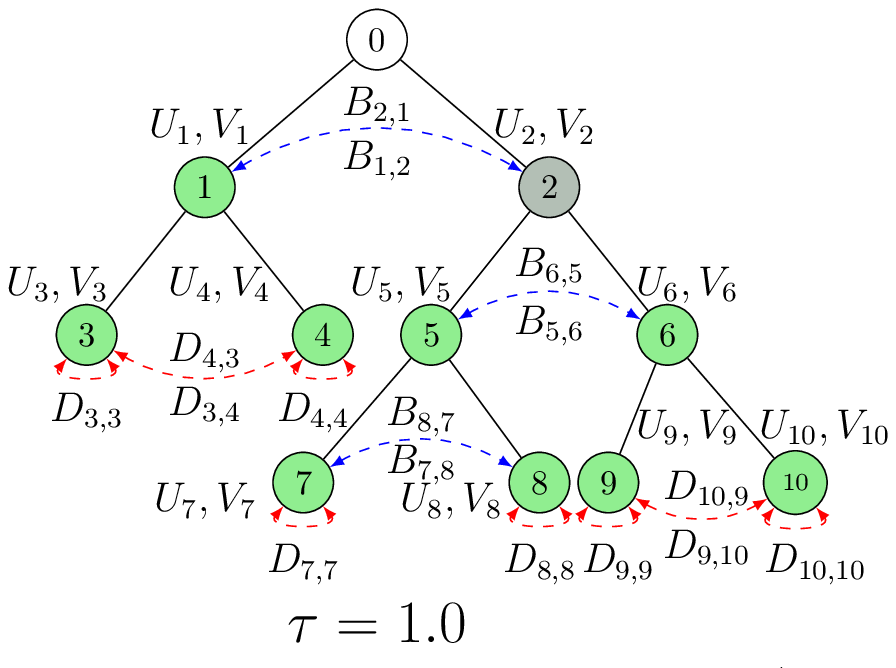}
 }
 & \subfloat[Conceptual diagram of approximation matrix.  Red submatrices are not approximated, blue ones are.]{
 \label{fig:hmatrix}
     \includegraphics[width=0.22\textwidth]{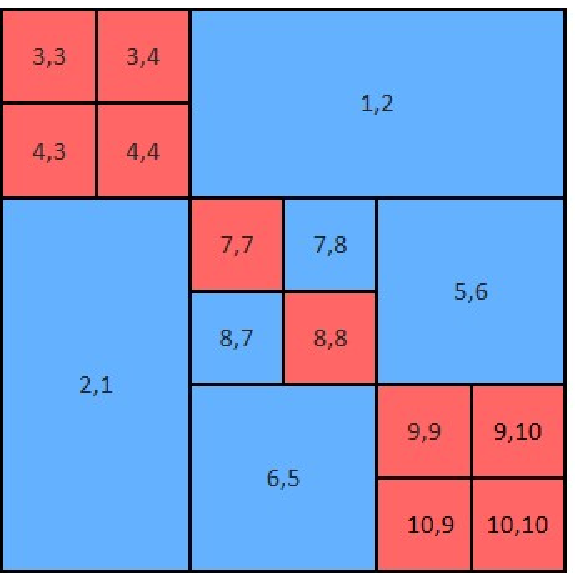}
     } \\
     \hline
 \subfloat[Library code for evaluation for $D$ and $V$ matrices]{
 \label{fig:libcode}
 \input{Figures/library-v1.tex}
 } 
 & 
 \subfloat[MatRox code for evaluation for $D$ and $V$ matrices. 
 ]{
 \label{fig:matcode}
 \input{Figures/matroxcode.tex}
 }
 & 
 \begin{tabular}{l}
      \subfloat[Structure sets]{
      \label{fig:sets}
      \includegraphics[width=0.32\textwidth]{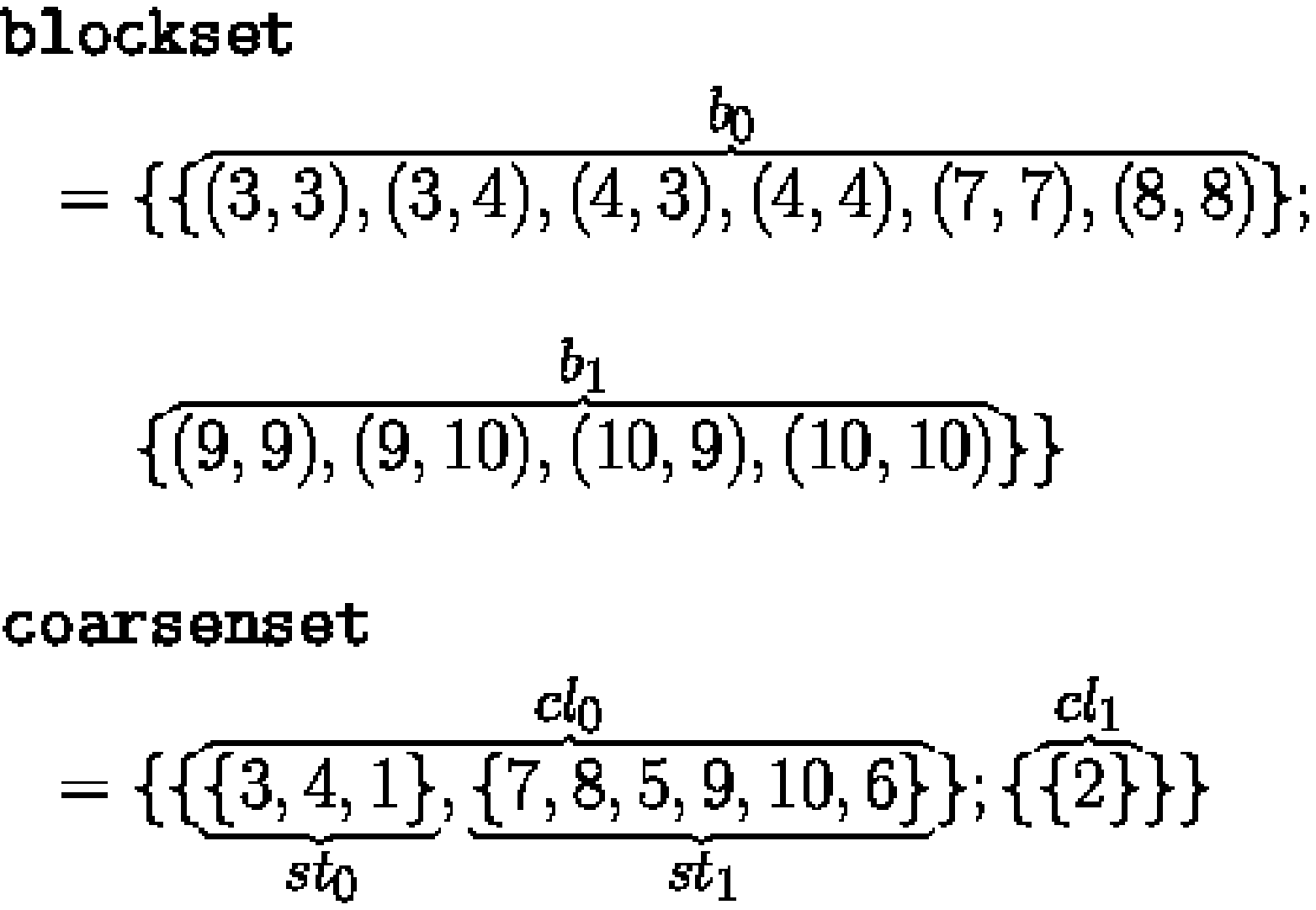}
      }  \\
      \subfloat[The $D$ generators stored in CDS]{
      \label{fig:dgen}
      \includegraphics[width=0.28\textwidth]{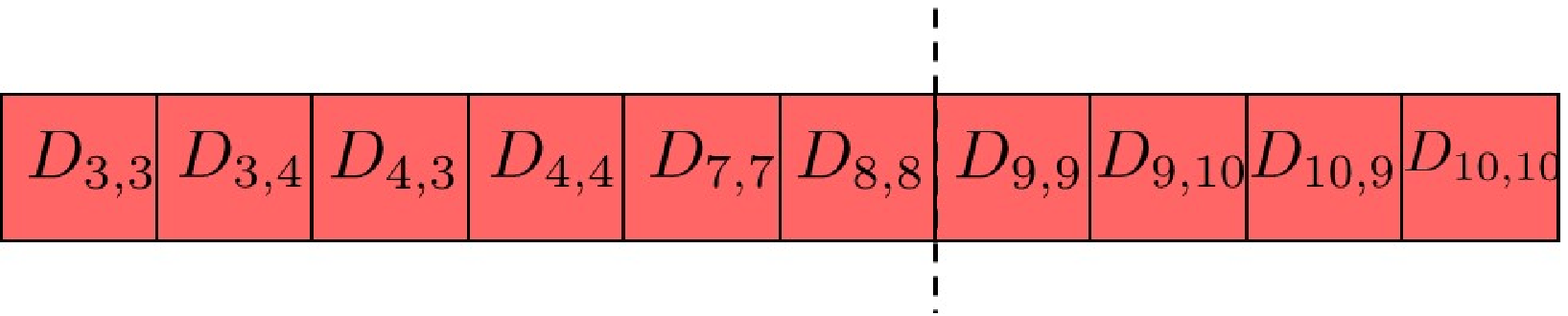}
      } \\
      \subfloat[The $V$ generators stored in CDS]{
      \label{fig:vegen}
      \includegraphics[width=0.28\textwidth]{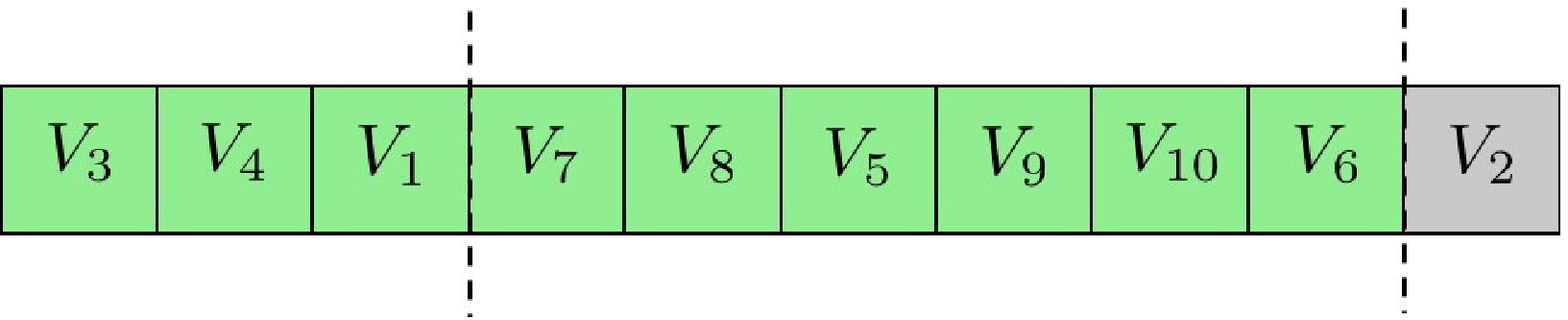}
      }
 \end{tabular}

\end{tabular}
\caption{Given a pointset  in Figure \ref{fig:points}, a kernel function, and other parameters, it is possible to approximate the $K$ matrix that results from applying the kernel function to  points.  Figure \ref{fig:htree} illustrates the hierarchical organization of subpartitions of the pointset and the submatrices that will be generated to approximate interactions between points in various partitions.  Figure \ref{fig:libcode} shows a typical library implementation of HMatrix evaluation using the CTree. Figure \ref{fig:matcode} is the implementation of HMatrix evaluation in MatRox.
 MatRox groups
the $D$ submatrices into blocks and
the $V$ matrices into coarsen level sets.}
\label{fig:motive}
\end{figure*}


\subsection{Background} 
Current library implementations of hierarchical matrix approximations typically have an interface as follows. The user provides to the library, the  pointset shown in Figure \ref{fig:points}, an admissibility parameter $\tau$, a kernel function, and a desired \textit{block approximation accuracy (bacc)}.
The compression phase approximates the kernel matrix, implicitly created using the points and the kernel function.
The admissibility parameter $\tau$ dictates how pairs of points are determined to be far or close, and the
block approximation accuracy \textit{bacc} indicates how closely submatrices need to be approximated.
A representation of the compressed matrix 
is the input to the evaluation code that multiplies the HMatrix 
with another matrix or vector. Compression is typically expensive, however, the objective is to compress the kernel matrix once and reuse over many evaluations, e.g. multiple matrix-vector multiplications or a matrix-matrix multiplication.


\textit{Compression.} To approximate $K$,  points are first clustered into hierarchically-organized sub-domains.  Figure \ref{fig:points} shows a clustering that creates 10 sub-domains.  Figure \ref{fig:htree} shows a cluster tree for this clustering with each of the tree nodes representing a sub-domain.  If two sub-domains $i$ and $j$ are \textit{Far} from each other, their interaction, ($i$,$j$), is approximated. Interactions between sub-domains \textit{Near} to each other are not approximated. The admissibility condition  $\tau dist(\alpha,\beta) > (diam(\alpha) + diam(\beta))$ in which $dist(\alpha,\beta)$ is the geometrical distance between the two sub-domains $\alpha$ and $\beta$, $diam(\alpha)$ is the diameter of $\alpha$, and  $\tau$ is the input admissibility parameter, determines near-far interactions.   The added dashed edges  in blue and red to the cluster tree in Figure \ref{fig:htree} represent these interactions and together form the HTree. The red edges are near interactions  and blue edges show the far interactions.

Figure \ref{fig:hmatrix} shows an example approximated ${K}$ matrix  with colored sub-blocks.  The blue-colored blocks are matrix blocks that are approximated during the compression phase\footnote{We use interpolative decomposition \cite{martinsson2011randomized} for approximations.}. The degree to which each block is approximated is determined by the \textit{submatrix-rank} (srank).  The srank is adaptively tuned to meet the user-requested  \textit{block approximation accuracy (bacc)}. The red  blocks are not approximated.


\textit{Evaluation.} 
%
The approximated matrix  $\widetilde{K}$ is never explicitly assembled, instead the HTree is used during evaluation to compute the desired HMatrix-matrix multiplication.
Figure \ref{fig:libcode} shows a typical library implementation of the evaluation phase. 
Existing library implementations of evaluation  are classified into \textit{(i)} Loops with reduction that operate on the near and far interacting nodes in the HTree; \textit{(ii)}  Loops with sparse dependencies between parents and children that do a bottom-up or top-down, level-by-level traversal of the cluster tree to generate other portions of the approximated matrix. Lines \ref{lin:snear}-\ref{lin:fnear} in Figure \ref{fig:libcode} show the reduction loop computing near interactions by operating on the $D$ submatrices; operation on the $B$ submatrices is of the same loop type. Lines \ref{lin:stree}-\ref{lin:ftree} in Figure \ref{fig:libcode} show the loop over the CTree that operates on the $V$ matrices with bottom-up traversal; the $U$ submatrices are operated on with the same loop type but using top-down traversal.   Some library  implementations perform the  level-by-level traversal with a synchronization between levels and others place tasks into a dynamic task graph to enable run-time load balancing.

\subsection{Approach implemented in MatRox}  
\label{sbs:ME}
MatRox is composed of an inspector and an executor. The inspector is a modularized implementation of  compression. 
It analyzes structure to generate optimized evaluation code and  to store the submatrices associated with nodes in the cluster tree into an optimized data structure called Compressed Data-Sparse (CDS).  Together the optimization of the code and reorganization of the data lead to faster evaluation compared to libraries.  Additionally, when the inputs to the inspector do not change, the inspection can be conducted at compile-time.

The approach presented here improves data locality and reduces synchronization costs by grouping computations and associated data when the computations share data and are dependent on each other. For example, the computations involving the submatrices (9,9), (9,10), (10,9), and (10,10) are grouped together using \textit{blocking}. The  blocking algorithm analyzes the HTree to create a \textit{blockset} shown in Figure 1f that creates an order for computation.  This  enables the blocked loop in Figure \ref{fig:matcode} to be fully parallel, because the blocks are selected to eliminate reduction dependencies between block computations. Also, these submatrices  are stored next to each other in the CDS format to improve locality.    

The loop 
over the CTree is reorganized into coarsen levels and load-balanced sub-trees within those coarsen levels.   The \textit{coarsening} algorithm analyzes the cluster tree to create a \textit{coarsenset} shown in Figure 1f that contains the coarsen levels and sub-trees.  In Figure\ref{fig:htree}, there are two coarsen levels ($cl_0$,$cl_1$).  The green nodes are in coarsen level 0 and node 2 is in coarsen level 1 by itself (node 0 is not involved in any computation).  The coarsened loop in Figure \ref{fig:matcode} has a sequential loop over the coarsen levels and then a parallel loop over all of the sub-trees within that coarsenset.  The sub-trees are load-balanced based on the srank.  Sub-matrices associated with all of the nodes in a coarsenset are organized together in the CDS format  as shown in Figure \ref{fig:dgen} and \ref{fig:vegen}.  

An example of how MaRox is used is shown in Figure \ref{fig:input} in which the user provides the points, the kernel function, the admissibility condition, and block accuracy to the inspector. The output of the inspector is used by the executor to complete evaluation. 
The CDS stored submatrices, shown with $H$ in  Figure \ref{fig:input}, as well as the generated HMatrix-matrix multiplication code are used by the executor. 

In addition to the block and coarsen optimization, MatRox also specializes the evaluation code for a given matrix block. 
For example, since the parallelism in the HTree is less when we get closer the root, MatRox peels the last iteration of the nested computation to exploit block-level parallelism, e.g. with parallel BLAS.
With all these changes, MatRox obtains $9.06\times$ speedup compared to GEMM and $2.11\times$ compared to GOFMM for covtype dataset on Haswell. 

\begin{figure}
    \centering
    \input{Figures/matroxIn.tex}
    \centering

    \caption{How a user provides parameters and  calls the MatRox inspector for compression and executor for evaluation.}
    \label{fig:input}
\end{figure}

\section{Modular HMatrix Approximation}
\begin{figure*}
    \centering
    \includegraphics[width=2.0\columnwidth]{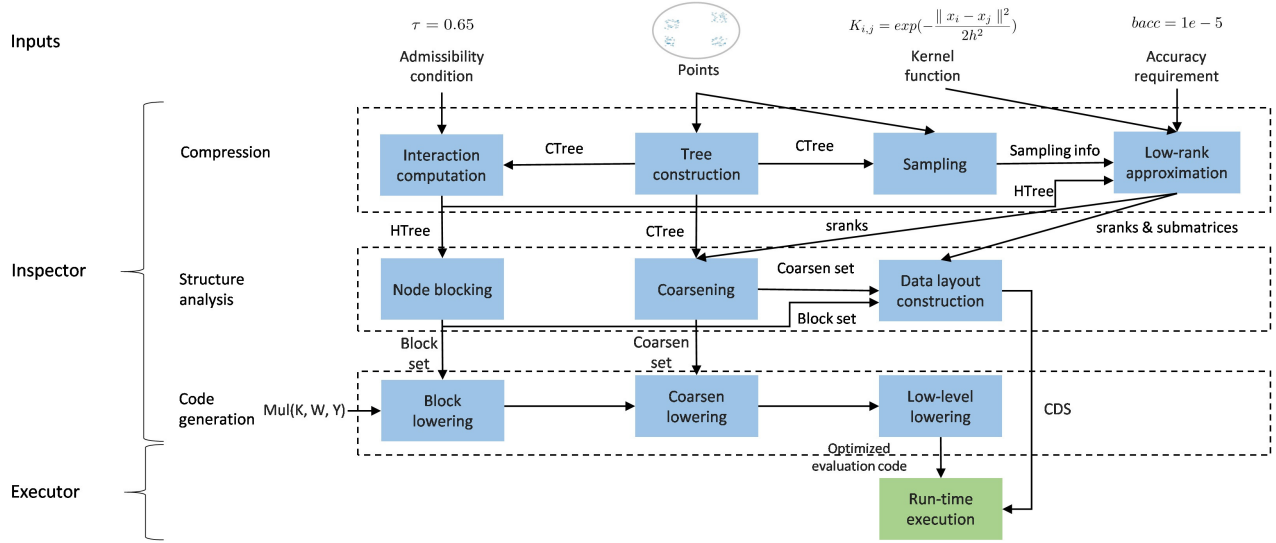}
    \caption{MatRox takes admissibility, points, kernel function, and accuracy as inputs and generates a storage format and an optimized code for HMatrix-matrix multiplication. It first compresses the matrix in the compression phase and then inspects the output of the compression phase in structure analysis. MatRox then uses the result of structure analysis, i.e. structure sets, to generate an optimized code and a storage format CDS. The MatRox executor runs the generated code with CDS. }
    \label{fig:structure}
\end{figure*}

MatRox consists of an inspector that generates specialized code and an efficient storage of the compressed data to improve locality and load balance in HMatrix-matrix multiplications. Figure \ref{fig:structure} shows the overview of MatRox. The inspector is separated in to three phases of modular compression, structure analysis, and code generation. The user-provided inputs are separately passed to their respective modules in compression. Compression generates the submatrices, sranks, as well as the CTree and HTree to be used by different components of  structure analysis. Information from structure analysis, i.e. the \textit{structure sets}, are used along with an internal representation of the HMatrix-matrix multiplication for  code lowering and specialization in the code generation stage. Finally the generated code and the submatrices stored in CDS are used by the executor for an efficient HMatrix-matrix multiplication.    This section describes the MatRox internals.

\subsection{Modularizing compression}\label{sec:sa}

MatRox provides a modularized design for the compression phase by defining four well-separated modules, i.e. interaction computation, tree construction, sampling, and low-rank approximation. Each module has well-defined inputs and creates and stores one or more of the outputs HTree, CTree, the \textit{sranks}, and the submatrices, which we  call the \textit{structure information}.
By modularizing compression we divide it into smaller pieces, each of which will take only the required user-provided inputs and/or inputs from another piece. With the tree construction, interaction computation,  low-rank approximation, and  sampling modules, the structure information are separately stored and are passed to the structure analysis phase or to other parts in compression.  The following discusses each module in the compression phase.

\textbf{Tree construction and interaction computation.}
Poin-ts are inputs to the tree construction module and the output is the CTree. The CTree is  constructed recursively using a  partitioning algorithm with the  tree root as the entire pointset. The partitioning terminates when the number of points in the leaf node is less than a pre-defined constant \textit{m}, i.e. \textit{leaf size}. MatRox uses two partitioning algorithms, kd-tree \cite{omohundro1989five} and two-means~\cite{rebrova2018study}, which are respectively used for low ($d \leq 3$) and high ($d >3 $) dimensional data. The interaction computation module takes as input the CTree and  the admissibility parameter to find near and far sub-domains using the admissibility condition. It then adds the interactions to the CTree to create  and store the HTree. 

\textbf{Low-rank approximation and sampling.}
The inputs to the low-rank approximation module are the HTree,  kernel function, block-accuracy, and the sampling information and the outputs are the sranks  and the submatrices. Interpolative Decomposition (ID)  \cite{martinsson2011randomized}  is used to create the $U$, $V$, and $B$ submatrices with low-rank approximation and the full-rank $D$ submatrices, i.e. the near blocks, are stored without approximation.
Each low-rank block in that is compressed with a rank that is adaptively tuned to meet the input block-accuracy specified by the user. The rank with which a block is approximated with is stored in the sranks vector. ID can be expensive for larger block sizes \cite{march2017far}, thus, sampling techniques are used to reduce the overhead of ID \cite{march2016askit}.

Sampling is a separate module in MatRox  compression  that takes only the points and the CTree as inputs and generates the sampling information for each sub-domain to be used by the low-rank approximation module. MatRox uses nearest-neighbour sampling~\cite{march2015askit} to  reduce the overhead of low-rank approximation. The sampling module first takes the unclustered points to generate the \textit{k}-nearest-neighbour list for each point \cite{dasgupta2008random}.  \textit{k} is the number of sampled points, \textit{sampling size} \cite{yu2017geometry}, and is a predefined constant.  Finding the exact k-nearest-neighbours of all points can be costly (points with high dimensions).  To reduce this overhead,  we use a greedy search based on random projection trees that recursively partitions the points along a random direction \cite{dasgupta2008random}.  The  lists are then combined for each block using the clustering in CTree to form a nearest-neighbour list for the corresponding sub-domain/block. Finally,  importance sampling \cite{march2016askit} selects from the nearest-neighbour list of a block and generates the sampling information for that block.

\subsection{Structure analysis}\label{sec:sa}
As shown in Figure~\ref{fig:structure}, after finishing compression, all structure information is known, and MatRox analyzes this information using the blocking and coarsening algorithms to create the coarsenset and blockset that are later used to generate specialize code for HMatrix-matrix multiplication. The submatrices and sranks from the modular compression phase are used with the sets to store the HMatrix in the  CDS format. In this section, we describe this structure analysis. 

\textbf{Blocking.} As shown in Algorithm \ref{alg:nodeblock}, the blocking algorithm takes the $HTree$ and an additional parameter called  \textit{blocksize}, as inputs and creates the \texttt{blockset}. We only show the blocking algorithm for near interactions; far interactions follow a similar algorithm. The blocking algorithm in lines \ref{line:bnblk}-\ref{line:snblk}, maps a near interaction between nodes 
$i$ and $j$ to the location of ($i$/blocksize, $j$/blocksize) in the \texttt{blocks} array. This mapping increases the probably that interactions that involve the same node are in the same block which increases locality. However, as shown in line 5 of Figure 1d all interactions ($i$,$j$) will write  to the location $i$ of \texttt{y}, thus these interactions have to be put in the same block of \texttt{blockset} to eliminate synchronization; this is implemented in Lines 10-16 of Algorithm \ref{alg:nodeblock}. 

\begin{algorithm} [!t]
\caption{Blocking for near interactions}	\label{alg:nodeblock}
	\SetAlgoLined
	\begin{small}
				\SetKwInOut{Algorithm}{Algorithm}
			\SetKwInOut{Input}{Input}
			\SetKwInOut{Output}{Output}
			\Input{$HTree$, blocksize}
			\Output{blockset}
			$blockDim$ = (HTree.numNodes - 1 + blocksize) / blocksize \\
			blocks[$blockDim$,$blockDim$] = {(0,0)}; \\
			\tcc{Find blocks based on near interactions}
			\For{every node i $\in$ HTree and i != root }{ \label{line:bnblk}
			    iid = ($i$-1) / blocksize \\
			    \For{node j $\in$ HTree.near[i]}{
		        jid = ($j$-1) / blocksize \\
			    blocks(iid,jid).append($i$,$j$)    
			    }
			}  \label{line:snblk}
			
			\tcc{Add blocks into blockset }
			\For{i=0; i<blockDim; i++}{ \label{line:bset}
			    \For{j=0; j<blockDim; j++}{
			        \If{
			        blocks(i,j).size() > 0 }{
			          blockset[$i$].append(blocks($i$,$j$))  
			        }
			    }
			}\label{line:sbset}
	\end{small}
\end{algorithm}

\textbf{Coarsening.} The coarsening algorithm creates a coarsenset that optimizes the loops over the CTree by improving locality  while maintaining load balance. The algorithm is an adaptation of the Load-Balanced level Coarsening (LBC) method \cite{cheshmi2018parsy} with the difference that here the algorithm is designed  for binary trees and a different cost model based on sranks is used to balance load. As shown in Algorithm \ref{alg:partAlg},  coarsening takes the CTree, the sranks, number of sub-trees $p$, and a tuning parameter $agg$ as inputs and  generates a coarsenset. In lines \ref{line:bcoarsen}-\ref{line:ecoarsen}  the levels of the CTree are coarsened to build the coarsened levels. A level of a tree refers to nodes with the same height. \textit{Tree[lb:ub]} shows a coarsen level that includes nodes with levels in the range of lb-ub. Algorithm \ref{alg:partAlg} builds all disjoint trees in a coarsen level, line \ref{line:subtrees}, and stores them in \texttt{coarsenset} in post-order. For example in Figure \ref{fig:htree}, the disjoint trees of HTree[0:1] are sub-trees with a root node in 1, 5, and 6. This ensures all nodes with dependency are assigned to the same thread to improve locality. The coarsening algorithm computes the cost of each sub-tree using sranks in lines \ref{line:bcost}-\ref{line:ecost}. The subtree cost is related to the size of submatrices associated with the subtree nodes  and is determined by sranks. The computed costs are used in lines \ref{line:bpack}-\ref{line:epack} of Algorithm \ref{alg:partAlg} to merge the initial disjoint sub-trees with a  first-fit bin-packing algorithm~\cite{coffman1978application} and to create $p$ new sub-trees that are load balanced. These sub-trees will execute in parallel.

\textbf{Data layout construction.} In the final phase of structure analysis, MatRox uses the structure sets to store the generated submatrices in a format, which we call the compressed data-sparse (CDS),  that improves locality during the HMatrix-matrix multiplication. CDS follows the order of computations in the blocked and coarsened loops which is obtained from the structure sets. More specifically, the $U$, $V$ submatrices are stored in the order specified by the coarsenset and the $B$, $D$ submatrices are stored by the order specified by the near and far blocksets. The size of each submatrix is known with sranks and is used as the offsets in CDS (see
Figure~\ref{fig:sets} for an example).

\subsection{Code generation}\label{sec:codelowering}
Code generation in MatRox uses structure information to lower an internally generated abstract syntax tree (AST) to an optimized evaluation code. \autoref{fig:structure} shows different components of code generation. 
MatRox lowers the AST in either the block or the coarsen lowering stages or both.  The resulting lowered code from these stages iterates over the structure set.  Figure \ref{fig:matcode} shows an example lowered code where the blocked loop iterates over the \texttt{blockset} and the coarsen loop goes over \texttt{coarsenset}. The number of blocks and number of levels are used to determine whether the block and/or coarsen lowering  should be applied. If the number of blocks are larger than an architecture-related threshold, \textit{block-threshold}, MatRox applies block lowering. Similarly the number of levels and a \textit{coarsen-threshold}  is used to determine the application of coarsen lowering. These thresholds are defined to ensure there are enough parallel workloads that amortize the initial cost of launching threads. 
MatRox further optimizes the lowered code with low-level optimizations, using structure information, as shown in Figure~\ref{fig:structure}. 

\begin{algorithm}[!t]
		\caption{Coarsening }
		\SetAlgoLined

	\begin{small}
			\label{alg:partAlg}
			
			\SetKwInOut{Algorithm}{Algorithm}
			\SetKwInOut{Input}{Input}
			\SetKwInOut{Output}{Output}
			
			\Input{$CTree, p, agg, sranks$}
			\Output{coarsenset}

			
			$l = \lceil CTree.height / agg \rceil$ \\
			\For{ i=0; i<$l$; i++}{\label{line:bcoarsen}
	    		lb = $i$*agg; \\
				ub = ($i$+1)*agg; \\
				cl = disjoint\_subtrees($CTree$[lb:ub]); \label{line:subtrees} \\
			    coarsenset.append(cl);
			}       \label{line:ecoarsen}

			\tcc{Cost estimation for each node}
			\For{ node $ x \in CTree$}{\label{line:bcost}
				\uIf{ x $\in$ CTree.leafnodes}{
					$CTree$[$x$].cost = cost(sranks($x$)) \\
				}
				\Else{
					$CTree$[$x$].cost = cost(sranks($x$), srank(lchild($x$))+sranks(rchild($x$)))
				}	
			} \label{line:ecost}
			\tcc{Merge subtrees in each coarsen level}
			\For{i=0; i<$l$; i++ }{\label{line:bpack}
			    cl = coarsenset[$i$] \\
			    nPart = cl.size() $>$ $p$ ? $p$ : cl.size()/2 \\
			    coarsenSet[$i$] = bin\_pack(cl, nPart); \label{line:binpack}
			} \label{line:epack}
			
			
	\end{small}
\end{algorithm}
\section{Experimental Results}\label{sec:results}
We compare the inspector and executor performance of MatRox to the corresponding parts from STRUMPACK \cite{ghysels2016efficient}, GOFMM \cite{yu2017geometry}, and SMASH \cite{cai2018smash}, which are well-known libraries for HMatrix approximation.  
The inspector performance for MatRox is quite similar to the existing libraries.  The resulting matrix-matrix multiply performed by the executor is much improved over existing library implementations due to improvements in data locality and parallelism.

\subsection{Methodology}
We select a set of datasets, i.e. points, used also in prior work and shown in Table \ref{tab:dataset} from real-world machine learning  and scientific applications. Problem IDs 1-8 are machine learning datasets from the UCI repository \cite{bache2013uci} and are high-dimensional points. Problem IDs 9-13 are scientific computing points that are low-dimensional 
\cite{cai2018smash}. STRUMPACK only runs for small datasets, i.e. problem IDs 5, 6, 8, 13. We use a Gaussian kernel \cite{williams1996gaussian} with bandwidth of 5 when comparing to GOFMM and STRUMPACK. 
For comparisons to SMASH we use their default settings of kernel function $(1/\parallel x-y \parallel)$ and the scientific pointsets ID 9-13 (SMASH only supports 1-3 dimensional points); MatRox uses the same setting when compared to SMASH. The HMatrix is multiplied with a randomly generated dense matrix $W$ of size ${N\times Q}$ . 

\makeatletter
\newcommand{\thickhline}{%
    \noalign {\ifnum 0=`}\fi \hrule height 2pt
    \futurelet \reserved@a \@xhline
}
\newcolumntype{"}{@{\hskip\tabcolsep\vrule width 1pt\hskip\tabcolsep}}
\makeatother

Testbed architectures are Haswell with Xeon\texttrademark{} E5-2680v3,  12 cores, 2.5 GHz, 30MB L3, and KNL with Xeon\texttrademark{} Phi 7250, 68 cores, 1.4 GHz, and 34MB L3.    All tools are compiled with  icc/icpc 18.0.1 with \texttt{-O3}. 
For BLAS/LAPACK routines we use MKL \cite{wang2014intel}.
MatRox is implemented in C++ in double precision. The median of 5 executions is reported for each experiment.

When comparing to libraries we use their default settings and use the same in MatRox, e.g \textit{sampling size=32}, 
\textit{maximum rank=256}. MatRox-specific parameters are $agg=2$, $p=$  number of physical cores,  $blocksize=2$ for near and $blocksize=4$ for far interactions, \textit{coarsen-threshold=4}, and \textit{block-threshold=}  number of leaf nodes. $bacc$ is set to $1e-5$ for all experiments with MatRox and the libraries and the overall accuracy, i.e. accuracy of the HMatrix-matrix multiplication, is the same in MatRox and the libraries.
We choose the admissibility condition to match the library's default setting. STRUMPACK only supports HMatrix structures with a very large admissibility condition in which all off-diagonal blocks are low-rank approximated; also known as HSS. GOFMM uses a \textit{budget} parameter instead of admissibility,  which we also implement in MatRox. Recommended budget settings in GOFMM are 0.03 and 0, in our results we refer to the former $\mathcal{H}^2$-b and the later as HSS as its structure is HSS.  SMASH's default admissibility is 0.65, which we also use.

\subsection{Performance of the MatRox inspector}

\begin{figure}
    \centering
    \includegraphics[width=\columnwidth]{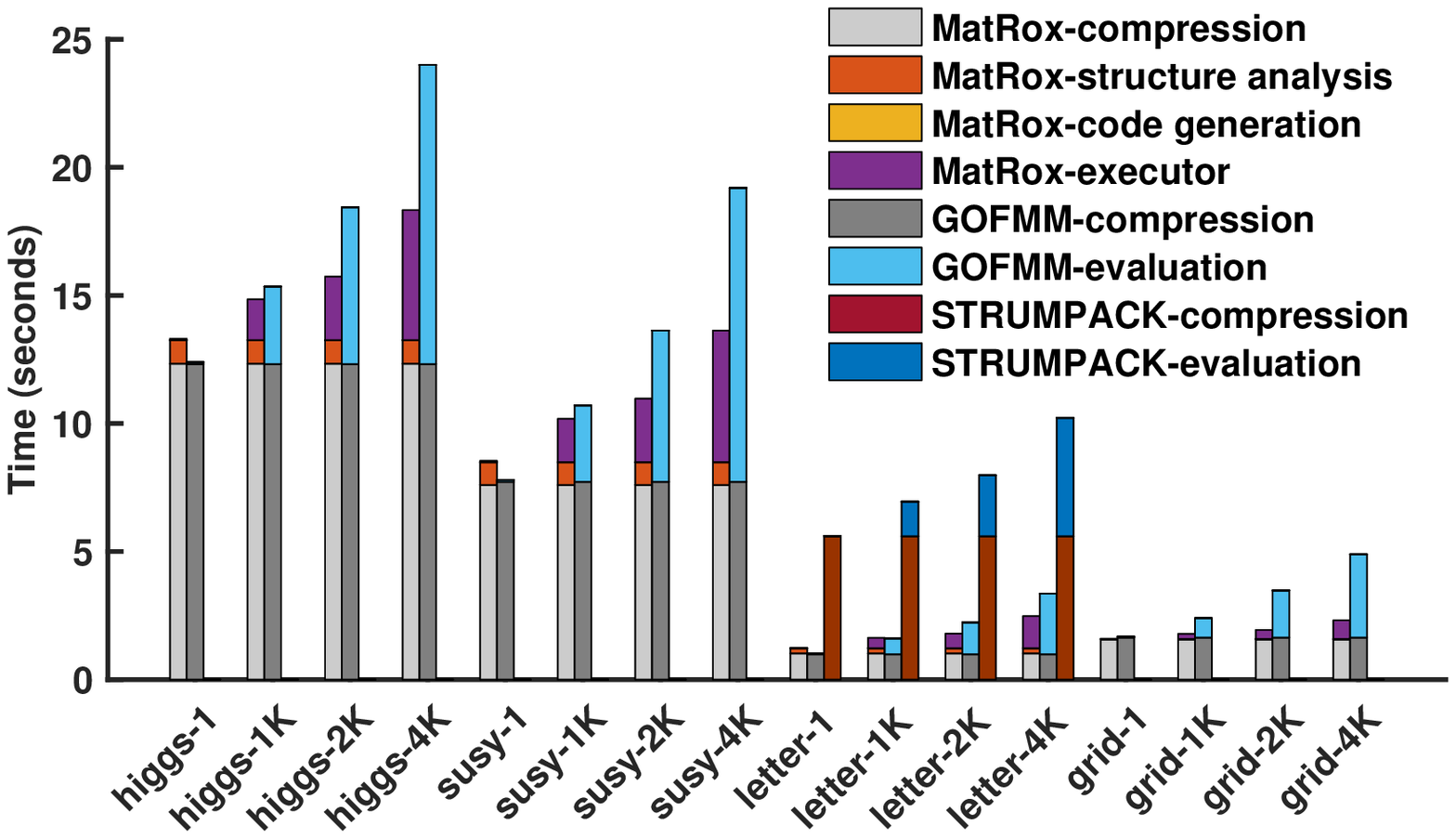}
    \includegraphics[width=\columnwidth]{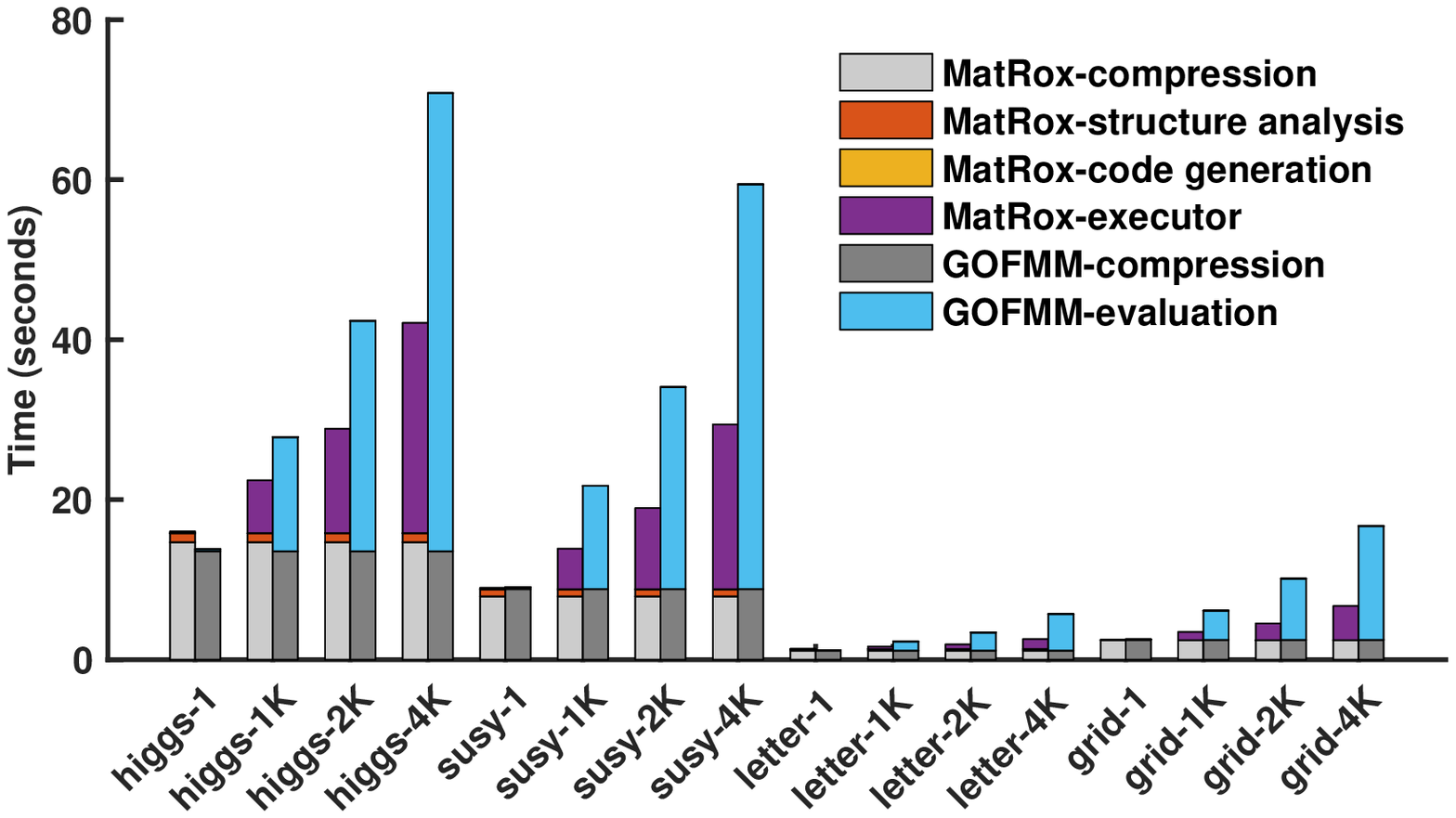}
    \caption{ The overall time in MatRox, GOFMM, and STRUMPACK for different values of $Q$ i.e., $1$, $1K$, $2K$ and $4K$ for HSS (top) and $\mathcal{H}^2$-b (bottom) on Haswell. Missing bars for STRUMPACK mean it could not run that dataset.
    }
    \label{fig:nrhs}
\end{figure}

MatRox's inspector time is similar to that of libraries since the time for structure analysis and code generation are negligible compared to the compression time as shown in 
Figure \ref{fig:nrhs}.   Structure analysis and code generation in MatRox is on average $8.1$ percentage of inspection time.
 The compression time of STRUMPACK is slower than MatRox and GOFMM because of using a different compression method. 

\begin{table}[]
    \caption{Data set: N is number of points, d is point dimension. }
    \centering
    \begin{small}
	\begin{tabular}{|m{5mm}|m{9mm} m{5mm} m{8mm} m{4mm} m{11mm} m{5mm} m{9mm}|}
		\hline 
		\textbf{ID}	& \textbf{1}& \textbf{2} & \textbf{3} &\textbf{ 4} & \textbf{5} & \textbf{6} & \textbf{7}  \\ 
		\hline 
		{Data} & covtype   & higgs & mnist & susy & letter & pen & hepmass \\  
		\hline
	{N}	& 100k & 100k & 60k & 100k & 20k & 11k  & 100k \\
	{d}  & 54     & 28 & 780 & 18 & 16 & 16 & 28 \\
\thickhline
		\textbf{ID}	& \textbf{8} & \textbf{9} & \textbf{10}  & \textbf{ 11} & \textbf{12} & \textbf{13} &  \\ 
		\hline 
		{Data}	& gas & grid  &random & dino  & sunflower  & unit &  \\ 
		\hline 
	{N}	& 14k & 102k & 66k & 80k & 80k & 32k  &  \\ 
	{d}	& 129 & 2   & 2  & 3 & 2 & 2 & \\
		\hline 
	\end{tabular} 
\end{small}
  
    \label{tab:dataset}
\end{table}

Figure \ref{fig:nrhs} also shows that the inspector time is amortized with increasing $Q$ (number of columns in matrix approximated K is multiplied by) because the evaluation time grows with $Q$.
The figure compares the MatRox  overall time, includes inspector and executor times, with  the overall time of libraries, i.e. compression and evaluation, for four  datasets and  $Q$ sizes of 1, 1K, 2K, and 4K on Haswell.
The compression time for both $\mathcal{H}^2$-b and HSS and for all tools will not change for $Q=1$ and a larger $Q$.
For example, for susy with $\mathcal{H}^2$-b, MatRox's overall Speedup vs GOFMM is $1.56\times$ for $Q=1K$ and $2.02\times$ for $Q=4K$.  Figure \ref{fig:nrhs} does not include SMASH because SMASH only supports matrix-vector multiplication ($Q=1$). We compared MatRox with SMASH for $Q=1$ and  our results show that the overall time of MatRox and its evaluation time is on average 1.1$\times$ and 1.6 $\times$ faster. 

 The benefits of improving evaluation  with MatRox are  more for larger $Q$s. In  scientific and machine learning applications, $Q$ is typically large and often close to $N$, shown in Table~\ref{tab:dataset}. Examples include multigrid methods in which the coefficient matrix  is multiplied by a large matrix  \cite{briggs2000multigrid},  Schur complement methods in hybrid  solvers \cite{yamazaki2010techniques}, high-order finite-elements \cite{dong2014step}, as well as direct  solvers \cite{im2004sparsity}.
 We also ran the un-approximated matrix-matrix multiplication $KW$ with GEMM. For the tested datasetes on average MatRox's overall time is 
 18$\times$ faster than GEMM for $Q=2K$; 
  the speedups obtained with HMatrix evaluation are significantly higher than GEMM for larger $Q$s. We use a $Q=2K$ for the rest of our experiments unless stated otherwise.

\subsection{The performance of the MatRox executor}
\begin{figure}
    \centering
    \includegraphics[width=\columnwidth]{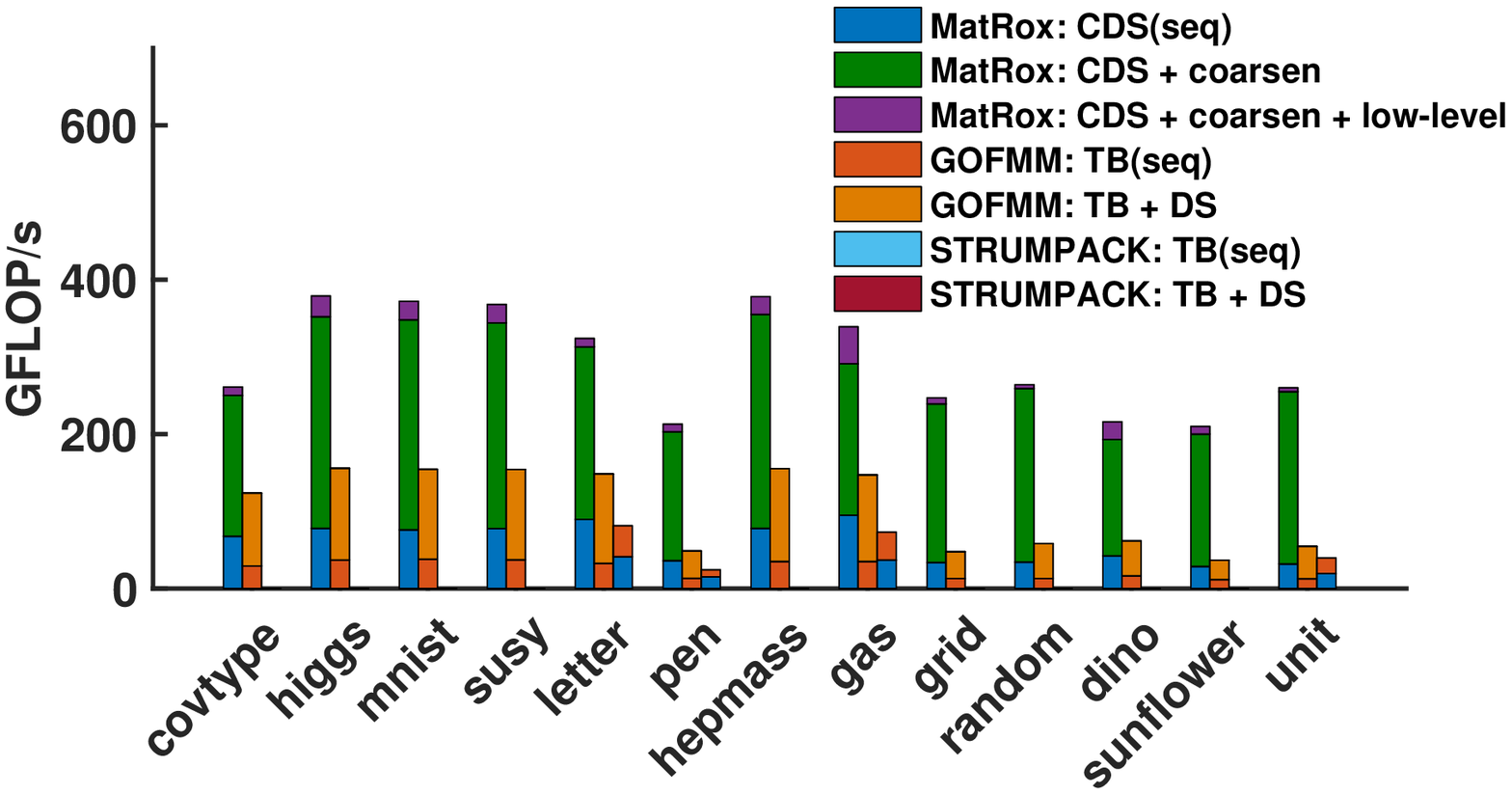}
   \includegraphics[width=\columnwidth]{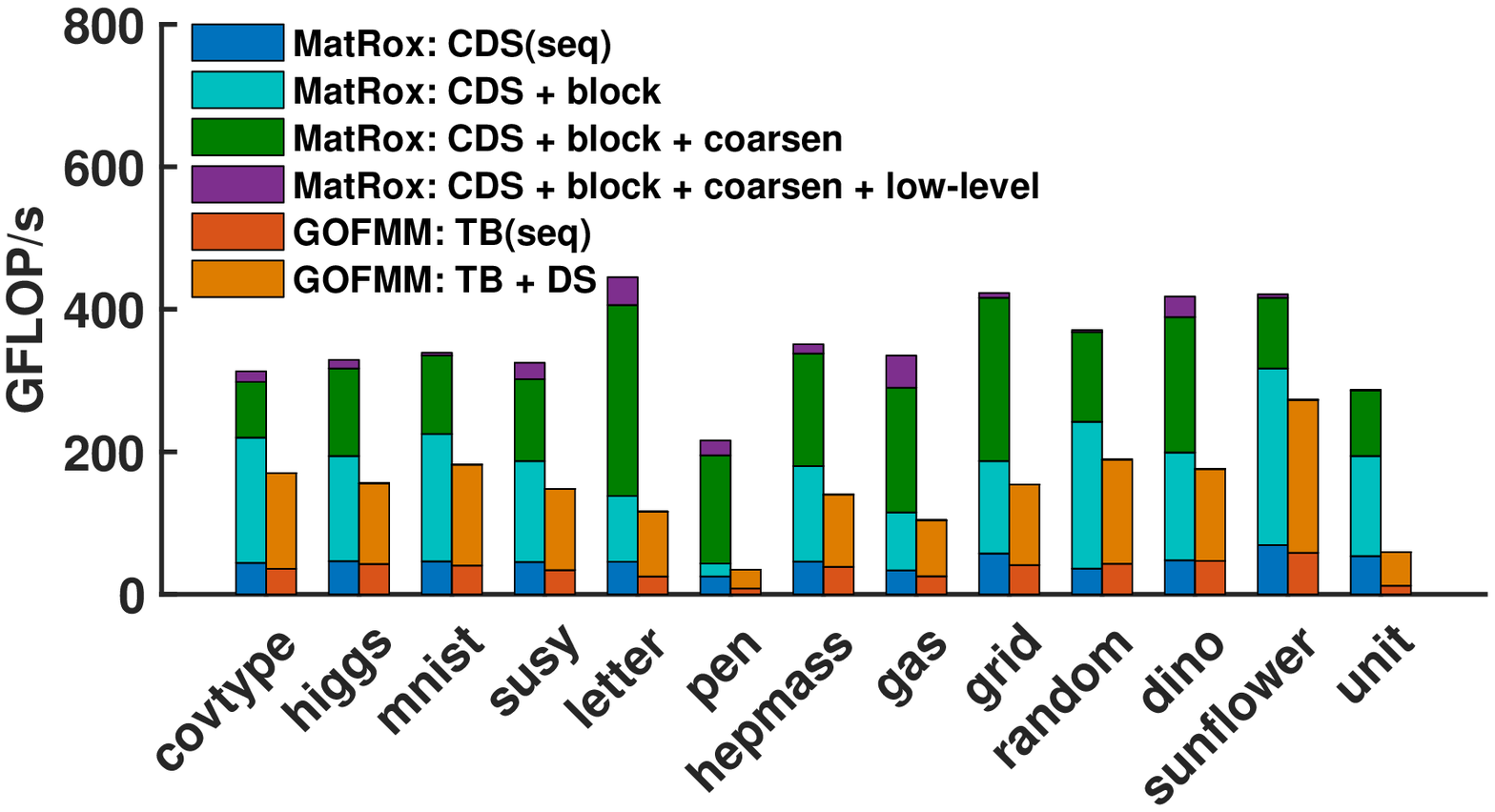}
    \caption{The performance of executor/evaluation in MatRox vs. GOFMM for HSS (top) and $\mathcal{H}^2$-b  (bottom) on Haswell. Labels seq, TB, and DS are sequential, tree-based format, and dynamic scheduling respectively. Effects of CDS, coarsening, blocking, and low-level transformations are shown separately. Missing bars for STRUMPACK mean it could not run that dataset.
    }
    \label{fig:sep}
\end{figure}

%
%
Figure \ref{fig:sep} shows the performance breakdown of the MatRox's executor (evaluation) versus GOFMM and STRUMPACK on Haswell.   As shown MatRox's evaluation time is on average 3.41$\times$ and 2.98$\times$ faster than GOFMM in order for HSS and $\mathcal{H}^2$-b and is on average 5.98$\times$ faster than STRUMPACK. The performance breakdown shows the effect of CDS as well as the block and coarsen algorithms. To show the effect of CDS, we run both the MatRox executor, that uses CDS, and GOFMM and STRUMPACK, that use a tree-based storage format,  with a single-thread and label in order with CDS (seq) and TB (seq). Coarsen, block, and low-level lowering applied in MatRox are labeled with coarsen, block, and low-level. The parallel version of GOFMM and STRUMPACK use a dynamic scheduling  labeled with DS in Figure \ref{fig:sep}. 



%
The different admissibility conditions in HSS and  $\mathcal{H}^2$-b  allows us to demonstrate  MatRox's performance for different HMatrix structures. HMatrix structures differ by the number of blocks that they low-rank approximate, which changes the ratio of loops over the CTree to loops with reduction. 
Because in HSS no off-diagonal blocks are full-rank, loops over the CTree dominate its execution time. 
%
As a result, from Figure \ref{fig:sep}, coarsening  contributes to a performance improvement of on average  $79.2\%$ for HSS, which is more that the  $46.8\%$ on average improvement from coarsening for $\mathcal{H}^2$-b. Also, while Blocking contributes on average $38.3\%$ to the performance of the MatRox generated code for $\mathcal{H}^2$-b on Haswell, block lowering is never activated for HSS since the number of loops with reduction, i.e. near-far interactions, never exceeds the \textit{block-threshold}. 
MatRox peels the last iteration of loops over the CTree.  Low-level transformations lead to on average $6.28\%$ and $4.24\%$ performance improvement of the MatRox code in HSS and $\mathcal{H}^2$-b respectively.  Since HSS is dominated by loops over the CTree, the effect of low-level transformation is also more prominent in HSS. 
\begin{figure}
    \centering
    \includegraphics[width=0.96\columnwidth]{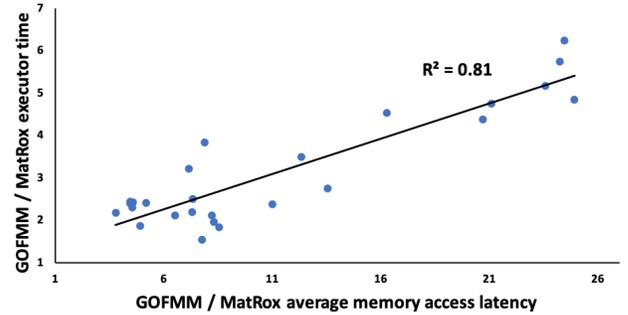}
    \caption{Effect of improving locality on MatRox's speedup vs. GOFMM on Haswell. Average memory access latency shows the average cost of accessing memory. 
    }
    \label{fig:loc}
\end{figure}

  
One major goal of inspection and lowering in MatRox is to improve locality. Figure ~\ref{fig:loc} shows the correlation between the performance of MatRox generated code versus the cost of Average memory access latency among all datasets for both HSS and $\mathcal{H}^2$-b on Haswell.
We measure the average memory access latency~\cite{hennessy2017computer} that is computed based on the number of memory accesses, miss-ratio of different cache levels, and TLB, and use it as a proxy for locality.
We use the \texttt{PAPI} \cite{terpstra2010collecting} library to collect L1, LLC (Last-level Cache), TLB access and misses and number of memory accesses. 
The coefficient of determination or R$^2$ is 0.81 that shows a good correlation between speedup and memory access latency. 

\subsection{Scalability of the MatRox executor}
\begin{figure}
    \centering
    \includegraphics[width=0.48\columnwidth]{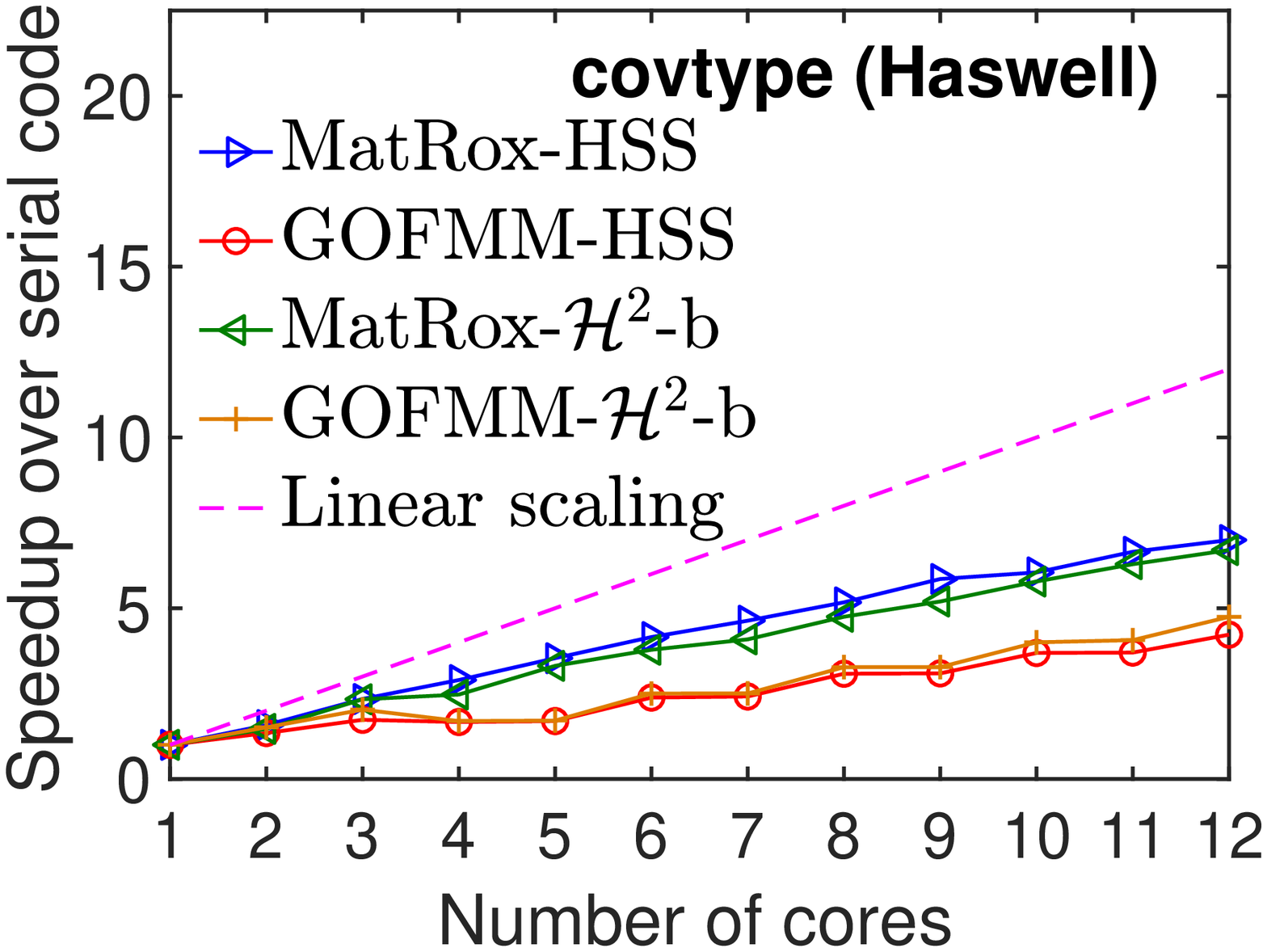}
    \includegraphics[width=0.48\columnwidth]{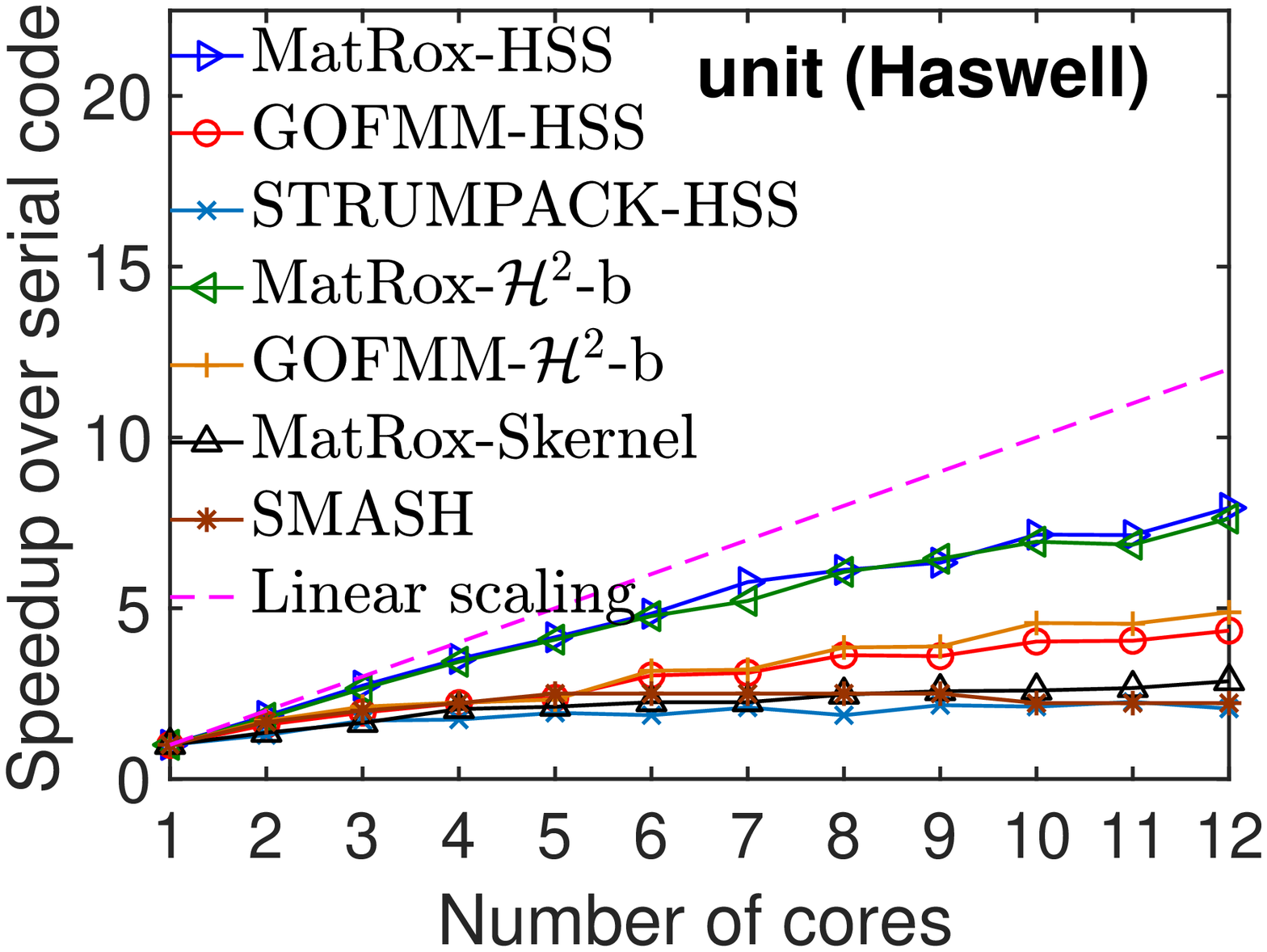}
     \includegraphics[width=0.48\columnwidth]{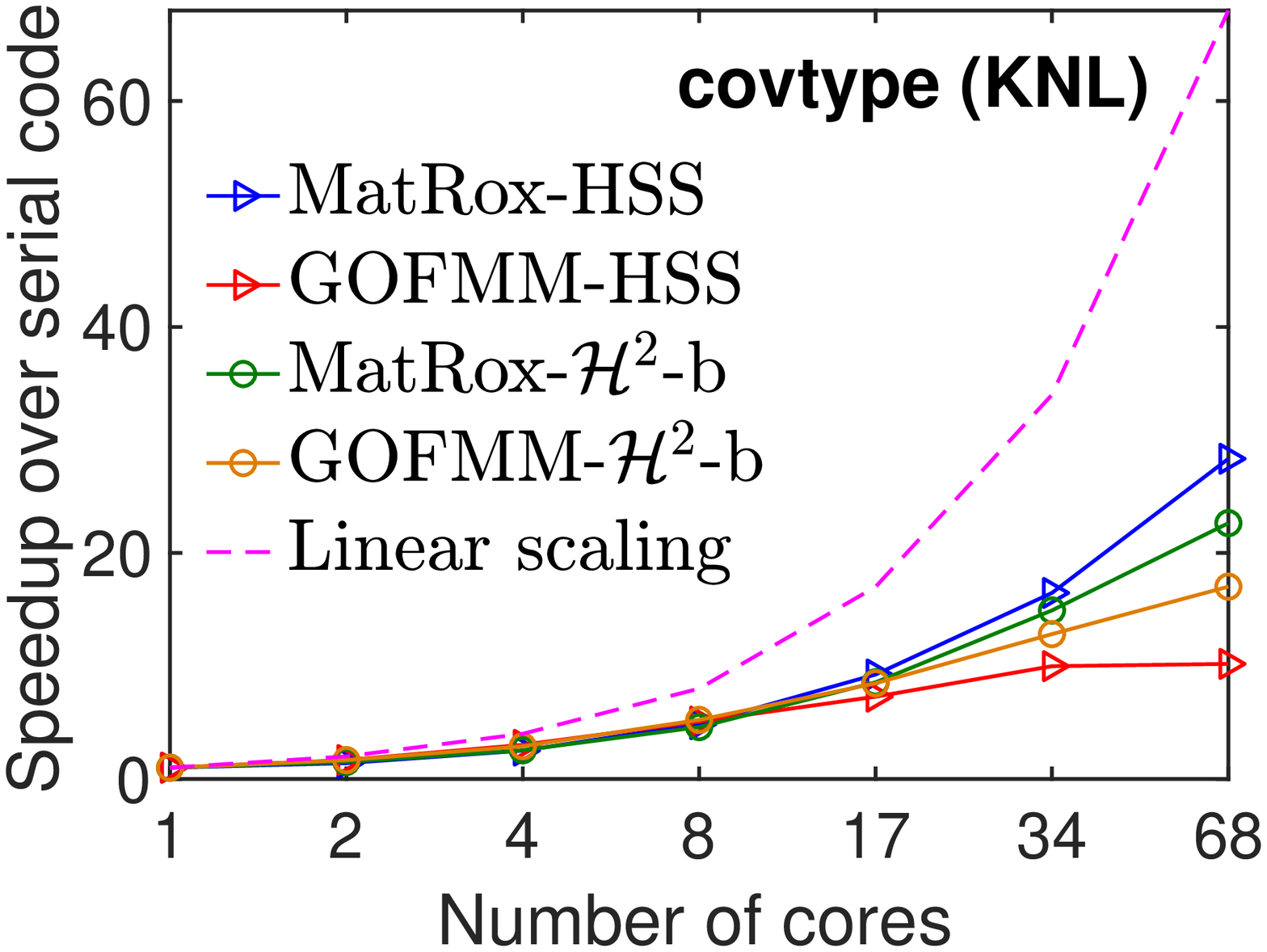}
    \includegraphics[width=0.48\columnwidth]{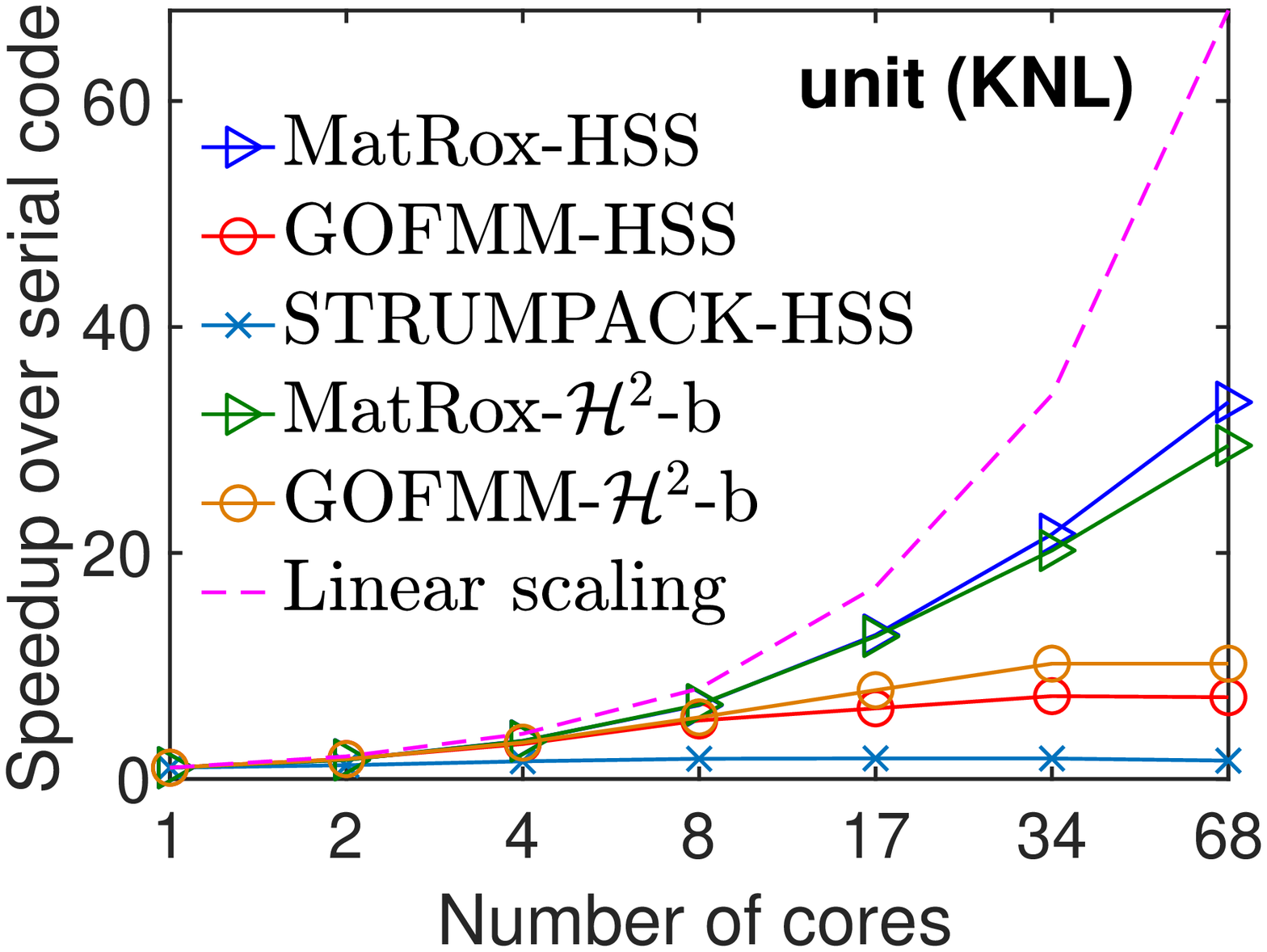}
    \caption{Scalability result on Haswell (top two) and KNL (bottom two) for datasets covtype (left) and unit (right).
    }
    \label{fig:scal}
\end{figure}

Figure \ref{fig:scal} shows the scalability  MatRox executor vs. GOFMM, STRUMPACK, and SMASH for two  datasets on Haswell and KNL; other datasets follow a similar trend. SMASH does not support\textit{ covtype} and in the figure, MatRox-Skernel is MatRox with SMASH settings.  
We select KNL in addition to Haswell to demonstrate MatRox's strong scalability on more cores, i.e. 68 cores of KNL. MatRox scales well on both architectures while the libraries show poor scalability with increasing number of cores. For example, GOFMM's performance reduces from  34 to 68 cores. MatRox's strong scaling is because coarsening and blocking  improve locality and reduce synchronization while maintaining load-balance. 


\section{Reusing Inspection}\label{sec:matroxrc}


\begin{figure}
    \centering
    \input{Figures/matroxru_in.tex}
    \caption{Reusing inspection in MatRox}
    \label{fig:matrucode}
\end{figure}

The modular design in MatRox enables the reuse of specific outputs of the inspector when parts of the input change. In libraries, any change to the inputs results in re-running the entire compression and evaluation phases. However, when the kernel function and/or the input accuracy change in MatRox, the modules and components in the inspector that do not rely on these inputs can execute only once and be reused. MatRox enables this reuse by separating the inspector into two phases, i.e. \textit{inspector-p1} and \textit{inspector-p2}. The inputs to inspector-p2 are the kernel function and the input accuracy and it is composed of the  low-rank approximation, coarsening, and data layout construction modules in Figure 3. The remaining parts of the inspector in Figure 3  belong to inspector-p1.  A change in the kernel function and/or the accuracy only requires inspector-p2 and the executor to be re-ran. Figure \ref{fig:matrucode} shows an example code that allows for the reuse of inspector-p1 when the $bacc$ changes.

In scientific and machine learning simulations, typically the input accuracy and the kernel function change more frequently than  the input points and the admissibility condition. The reuse of inspector-p1 in MatRox reduces the overhead of these changes. 
For example in finite-elements the discretization, i.e. points, is often fixed~\cite{grasedyck2008parallel}, in statistical learning the training samples, i.e. points,  are reused during  offline training~\cite{hofmann2008kernel}, and in N-body problems the CTree is only reconstructed during rebuild intervals~\cite{bedorf2012sparse,miki2017gothic}. The admissibility condition also often remains the same in simulations and is known by the domain practitioner as it depends on the problem structure. However, users often need to tune the parameters in a kernel function specially in machine learning simulations.  For example, the bandwidth $h$ in the Gaussian kernel \cite{williams1996gaussian} is typically tuned with cross-validation to avoid overfitting~\cite{morariu2009automatic}.
%
Also, often the practitioner needs to tune the input accuracy (bacc) because the \textit{overall accuracy} of the HMatrix-matrix multiplication is not sufficient, i.e. $bacc$ is correlated with the overall accuracy with a loose upper bound \cite{march2015robust}, or the user might decide to trade accuracy for faster evaluation (or vise-vesa) with re-compression.  

\begin{figure}
    \centering
    \includegraphics[width=\columnwidth]{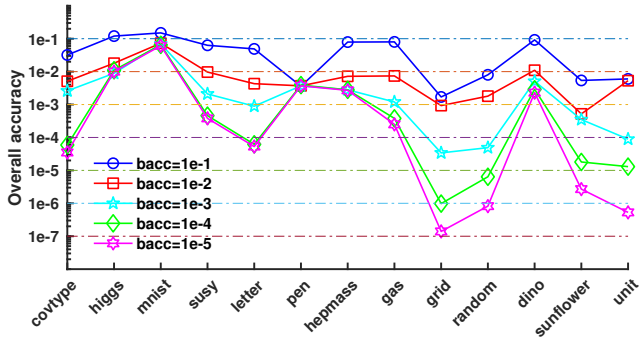}
    \caption{Input accuracy $bacc$ vs overall accuracy.}
    \label{fig:acc}
\end{figure}

\begin{figure}
    \centering
    \includegraphics[width=\columnwidth]{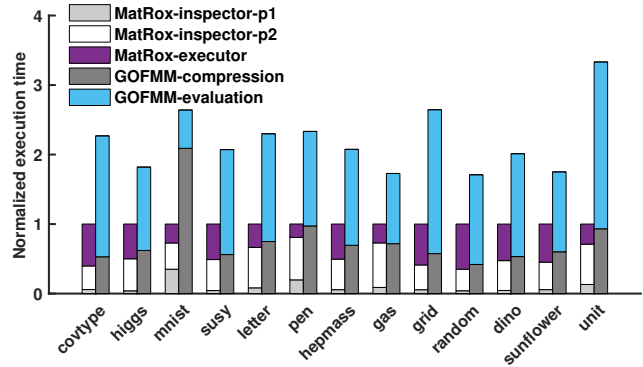}
    \caption{MatRox reusing inspection  for $\mathcal{H}^2$-b on Haswell.}
    \label{fig:matnrun}
\end{figure}

Figure \ref{fig:acc}, shows the correlation between  $bacc$ and overall accuracy $\epsilon_f$ obtained from $\epsilon_{f} = \lVert \widetilde{K}W - KW \rVert _{F} / \lVert KW \rVert_{F}$ for $\mathcal{H}^2$-b. As demonstrated, with a $bacc$ of $1\mathrm{e}{-3}$  more than 50\% of the datasets do not reach   an overall accuracy of $1\mathrm{e}{-3}$ and thus the user has to retune. The tuning becomes more important when more accurate results are required and also depends on the spectrum (i.e., eigenvalues) of the kernel Matrix.  
Figure \ref{fig:matnrun}   shows the MatRox's overall time compared to GOFMM for $\mathcal{H}^2$-b with 5 changes to the input accuracy \textit{bacc}, $1\mathrm{e}{-1}$ to $1\mathrm{e}{-5}$, with reusing inspector-p1. We do not include STRUMPACK for space but it follows a similar trend.   
As shown in the Figure,  MatRox's overall time is on average  $2.21\times$ faster than GOFMM. 
For high dimensional datasets  such as mnist sampling is expensive, $89.2\%$ of the compression time in mnist, and thus the reuse of sampling in inspector-p1 leads to a speedup of $2.64\times$ for minst compared to GOFMM.  MatRox with inspector reuse vs SMASH leads to on average  $1.37\times$ speedup with up to $2.4\times$ for sunflower. 


\section{Related Work}

The presented approach applies a modular inspector-executor strategy to HMatrix  Approximation.
This section summarizes previous approaches to improving the performance of HMatrix
approximation and related inspector-executor strategies that were used in other contexts.

\textbf{Hierarchical matrices. }
Hierarchical matrices are used to approximate matrix computations in almost linear complexity.  Hackbusch first introduced $\mathcal{H}$-matrices \cite{borm2003introduction,hackbusch1999sparse}, to generalize fast multipole methods \cite{greengard1987fast},  where the matrix is partitioned hierarchically with a cluster tree and then parts of the off-diagonal blocks are approximated. Later, $\mathcal{H}^2$-structures were introduced~\cite{hackbusch2000h2} which use nested basis matrices \cite{hackbusch2015hierarchical}, to further reduce the computational complexity of dense matrix computations. $\mathcal{H}^2$ has gained significant traction in recent years \cite{borm2003introduction,hackbusch2002data}. 
Hierarchical semi-separable (HSS) \cite{chandrasekaran2006fast,xi2016stability,xia2010fast}  are  a specific class of  $\mathcal{H}^2$ structures. MatRox supports HSS and other classes of $\mathcal{H}^2$  using a binary cluster tree; we abbreviate $\mathcal{H}^2$ matrix with HMatrix.

HMatrix approximations have a compression and an evaluation phase. Numerous algorithms have been studied for HMatrix compression ~\cite{chan1987rank, williams2001using,bebendorf2003adaptive} including  
interpolative decomposition (ID)~\cite{martinsson2011fast}. MatRox uses ID in its compression phase and contributes on improving the performance of the evaluation phase, which is the focus of many of the recent works on  HMatrix computations \cite{yu2017geometry, yu2018distributed}.

\textbf{Specialized libraries for HMatrix computations. } 
Numerous specialized libraries implement HMatrix evaluations on different platforms and for different evaluation operations.  HMatrix algorithms have been implemented on platforms ranging from shared memory~\cite{ghysels2016efficient,yu2017geometry,kriemann2005parallel}, distributed memory~\cite{march2015kernel,rouet2016distributed,march2015askit}, and many-core such as GPUs~\cite{march2015algebraic}. 
Hierarchical matrices have been studied to accelerate matrix factorization~\cite{xia2010fast,chandrasekaran2006fast,aminfar2016fast}. 
 Ghysels \textit{et. al.}~\cite{ghysels2017robust} introduces an algebraic preconditioner based on HSS.
Other work has improved matrix inversion \cite{martinsson2005fast} and matrix-vector/matrix multiplication \cite{chandrasekaran2006fast}.
STRUMPACK, GOFMM, and SMASH are the most well-known libraries that support HMatrix-matrix/vector multiplications. SMASH \cite{cai2018smash} supports 1-3D datasets while GOFMM \cite{yu2017geometry} and STRUMPACK \cite{ghysels2016efficient} also support datasets of higher dimension. MatRox generates code for HMatrix-matrix/vector multiplications for datasets of all dimensions on multicore platforms.   


\textbf{Inspector-executor approaches.} MatRox uses a domain-specific inspector-executor approach to generate code for HMatrix evaluation. Recent work ~\cite{strout2002combining,naumov2011parallel,govindarajan2013runtime,park2014sparsifying,liu2016synchronization} have proposed inspector-executors that inspect the data dependency graphs in sparse matrix computations to apply code optimizations that general compilers  cannot apply. 
Amongst them, inspectors based on  level-by-level wavefront parallelism \cite{venkat2016automating,rauchwerger1995run} are the most well-known, but do not optimize for locality and load-balance.   Cheshmi \textit{et. al.} \cite{cheshmi2018parsy} present an approach to improve wavefront inspectors, with the LBC algorithm by coarsening levels for better locality and creating balanced partitions. However, LBC only works for DAG from  a specific class of sparse matrix.  MatRox improves the data locality and parallelism in  reduction and tree-based loops for HMatrix approximation
with a novel coarsening method that uses a cost model of submatrix ranks and uses a specialized partition for binary trees.

\section{Conclusion}
We demonstrate  a novel structure analysis approach based on   modular compression to generate specialized code and an efficient storage that improves the performance  HMatrix approximations on multicore architectures. The proposed block and coarsen algorithms, improve locality in HMatrix evaluations while maintaining load-balance.  The modular approach used in MatRox, allows parts of the inspector to be reused when the  kernel function and  accuracy change.  MatRox  outperforms state-of-the-art  libraries for HMatrix-matrix multiplications on different multicore processors.

\bibliographystyle{ACM-Reference-Format}
\bibliography{sample-base,parsy-bib,matrox-arxiv-bib,Sympiler-bib}

\end{document}